\documentclass[preprint,12pt]{elsarticle}

\usepackage{amssymb}
\usepackage{graphicx}
\usepackage{subfigure}
\usepackage{amsmath}
\usepackage{esvect}
\usepackage{bm}
\usepackage{algorithm}
\usepackage{algorithmicx}
\usepackage{algpseudocode}
\usepackage{multirow}
\usepackage{array}

\journal{Knowl. Based Syst.}

\begin{document}

\begin{frontmatter}



\title{A Fast Recommendation Algorithm for Social Tagging Systems : A Delicious Case}


\author[a,b]{Yao-Dong Zhao}
\author[a,b]{Shi-Min Cai}
\ead{shimin.cai81@gmail.com}
\author[a,b]{Ming Tang}
\author[c]{Ming-Sheng Shang}
\address[a]{Web Sciences Center, School of Computer Science and Engineering, University of Electronic Science and Technology of China, Chengdu 610073, P. R. China}
\address[b]{Big Data Research Center, University of Electronic Science and Technology of China, Chengdu 610073, P. R. China}
\address[c]{Chongqing Institute of Green and Intelligent Technology, Chinese Academy of Sciences, Chongqing, 400714}

\begin{abstract}
The tripartite graph is one of the commonest topological structures in social tagging systems such as Delicious,
which has three types of nodes (i.e., users, URLs and tags). Traditional recommender systems developed based
on collaborative filtering for the social tagging systems bring very high demands on CPU time cost. In this paper,
to overcome this drawback, we propose a novel approach that extracts non-overlapping user clusters and corresponding
overlapping item clusters simultaneously through coarse clustering to accelerate the user-based
collaborative filtering and develop a fast recommendation algorithm for the social tagging systems. The experimental results show
that the proposed approach is able to dramatically reduce the processing time cost greater than $90\%$
and relatively enhance the accuracy in comparison with the ordinary user-based collaborative filtering algorithm.
\end{abstract}

\end{frontmatter}


\section{Introduction}
How users interact with the Internet has evolved from the birth of Web 2.0. The convenience of storing, publishing and sharing contents results
in an information overload for users when getting information which they are interested in.
The recommender system, with the purposes of improving user experiences and helping users to get information suited to their interests,
is one of the commonest modules in the web application. An increasingly influential set of websites
such as Delicious, Flickr, Youtube and LinkedIn provide the users with the service to tag the items such
as URL links, movies, photos, etc. The information given by tags reveals users' interests, depicts the items more precisely and provides more
opportunities and resources for data analysis and knowledge discovery. It is wisdom for us to exploit the abundant information of tags to recommend
interesting items to the users in the social tagging applications.

Although different applications of social tagging systems have different items, they all allow people to store and share the interesting contents and
can be modeled as tripartite graphs. Thus, the tripartite graph is one of the commonest topological structures in the social tagging systems which
have three types of nodes. In the social tagging applications, the nodes stand for users, items and tags, and users are interested in some kinds of
items which are attached with different tags. Figure 1 illustrates the modeled topological structure of the social tagging systems.

\begin{figure}[htbp]
\centering
    \includegraphics[width=7cm,height=4.8cm]{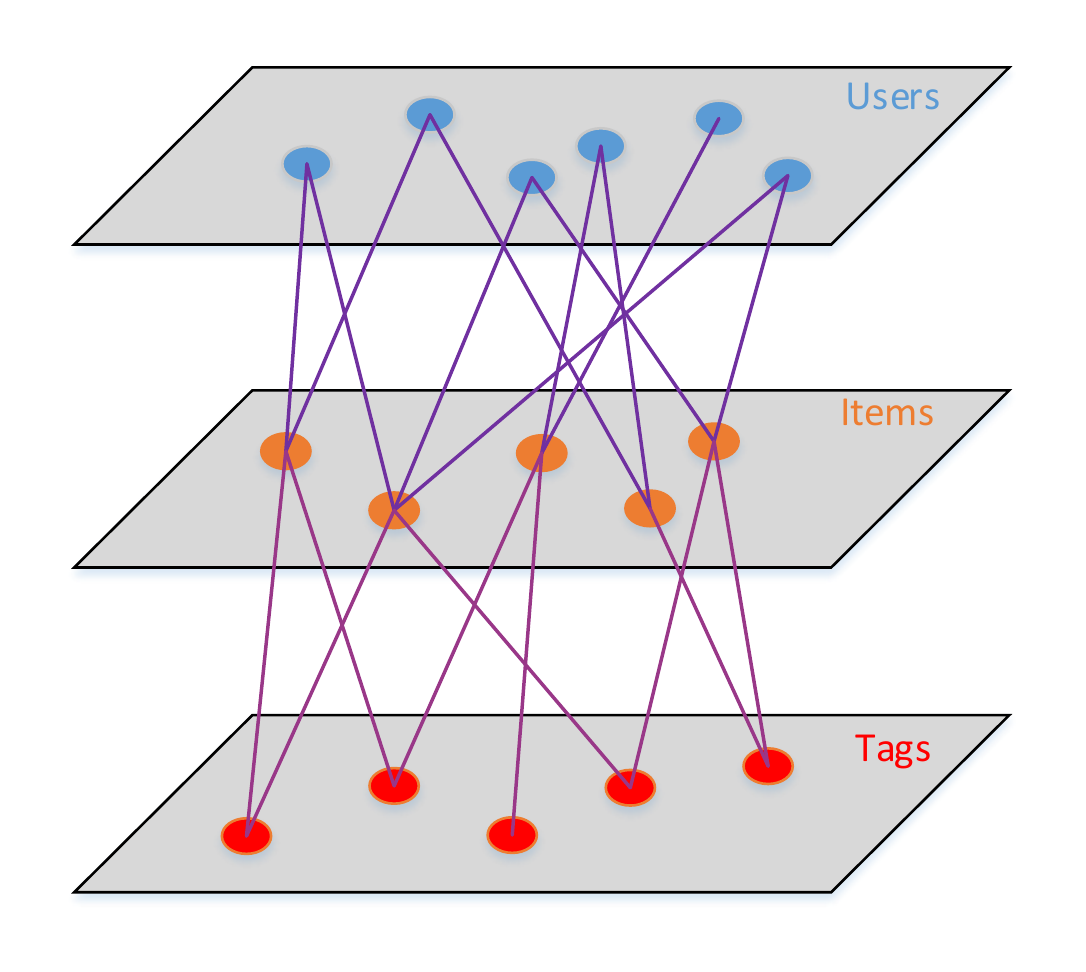}
    \caption{The structure of the social tagging systems.}
\end{figure}

The recommender systems have been well studied in the previous research in ~\cite{Admavicius.(2005),Aiolli.(2013),Bernardes.(2015),Shi(2014),Bobadilla(2013),Guan(2014),Zeng(2014)}.
At the same time, much effort has been made in the studies of social tagging systems in both structure domain and algorithm domain in ~\cite{Schenkel(2008),Ifada(2014),Ramage(2009),Huang(2014)}.
The main challenges of constructing a personalized recommender system for the social tagging applications are as follows:

\begin{itemize}
    \item \emph{Large volume of data}. In the online social tagging systems, there are enormous amount of users, items and tags. To analyse the data with the purpose of
    developing a recommender system to improve the user experiences brings high requirements on computational capabilities, and the time performance of the algorithm need to be outstanding.
    \item \emph{Diversity and novelty of items and tags}. There are various types of items and tags in the social tagging systems such as Delicious and Youtube.
    It is rarely easy to get the semantics of the items and tags. The challenges to understand the semantics may bring about noisy and inaccuracy models \cite{Zhou(2010),Vosecky.(2014)}.
    \item \emph{Timeliness}. The online applications produce enormous amount of data continuously. It is necessary to construct models which are fast enough to avoid being
    overwhelmed with tremendous new information. If the processing time of a model is too long, the results of the model can not be used because of the timeliness problem.
    \item \emph{Data sparseness}. In the real-world social tagging applications, there are users who have few store actions and share actions, moreover, many items are just
    be concerned few times. Considering this kind of data into the user modeling methods can result in inaccuracy models and slow down the algorithm~\cite{Vosecky.(2014),Lu.(2009)}.
    \item \emph{Cold start problem}. The cold start problem is the commonest challenge in recommender systems where recommendations to users are required when the users newly sign up
    in the applications. The cold start problems of different online applications have been studied in ~\cite{Martins(2013),Schein(2002),Mirbakhsh(2015)}.
\end{itemize}

In this paper, we address some of the above challenges (e.g., timeliness and data sparseness) by proposing a fast collaborative user model (FCUM)
which accelerates the ordinary user-based collaborative filtering (UCF) without accuracy loss. And we demonstrate
the performance of FCUM by an experimental evaluation on one real-world dataset which is crawled from the famous Delicious.
In the following paragraphs, we introduce the FCUM and acceleration method briefly, more details are in the following sections.

\textbf{Fast Collaborative User Model.}
The items stored, tagged and shared by users can provide rich information of the users' interests, moreover, the tags can also reveal the interests of the users.
In the FCUM, we exploit the information from both the items and the tags. Actually, if we conduct the recommendation procedure by using the information from all users,
the UCF model will be noisy and will bring about
high requirements on computational capabilities. Therefore, we extract non-overlapping user clusters and corresponding overlapping item clusters simultaneously and
construct the FCUM. In the experiment, we first get the scores of the items which stand for how much the users like the
items according to the behaviors of other users in the same cluster. Then a rank procedure is conducted, and we recommend the items to users in each cluster
according to the ranklist. Finally, we evaluate the FCUM in the aspects of accuracy and time performance. Moreover,
the conducted contrastive experiments will be depicted later.

\textbf{Acceleration Method.}
As mentioned above, we extract non-overlapping user clusters and corresponding overlapping item clusters simultaneously to construct the FCUM for recommending items to users.
In this way, we just need to conduct the UCF in the clusters separately that accelerates this ordinary algorithm. During this procedure,
we use a K-means-like approach to extract these clusters. First, the users are averagely and randomly distributed to clusters. Then, we update the centroid of each cluster
and redistribute the users in each cluster to clusters in terms of the similarities of users and the centroids of clusters. We just iterate this procedure less times and
will not wait for the convergence. Finally, the non-overlapping user clusters and corresponding overlapping item clusters are obtained respectively.
This coarse clustering procedure separates the useful information from the noise (i.e., redundant information) in the FCUM and accelerates the following UCF.

The rest of the paper is organized as follows: Section 2 gives out the details of the FCUM which is the foundation of the fast recommendation algorithm.
And, section 3 presents the design of our experiments on the Delicious dataset. Then, section 4 shows the results of the experiments and the evaluation indicators of the FCUM from many aspects. Finally, in Sec. 5, we conclude our findings and give out what we will do in the future.

\section{Fast Collaborative User Model}
Collaborative filtering has successfully been applied in recommender systems in \cite{Admavicius.(2005),Bernardes.(2015),Shi(2014)}. In this subsection,
we will comprehensively introduce the FCUM derived from collaborative filtering, which is an important module of the recommender system for a
social tagging application.

A social tagging system has various of users, items and tags. The behaviors of the users, including resource usages and annotation actions in the applications,
can be represented as the user-item-tag triple form. The recommender systems for the social tagging applications have been studied in ~\cite{Chelmis(2013),Peng(2010),Zhang(2010)}.
The main idea of this paper is to construct a FCUM that is \emph{fast} and \emph{accurate} enough for the recommendations in social tagging applications.
In the following paragraphs, we will give out the basic notations used in this paper.

As shown in Fig. 1, the social tagging system is denoted as a tripartite graph,
\begin{equation}
    G_{urt} = (U,R,T,E_{ur},E_{rt},E_{ut})
    \label{eq:1}
\end{equation}
where $U$, $R$, $T$ stand for the finite sets of users, items and tags, and $E_{ur}$ ,$E_{rt}$ and $E_{ut}$
(which is projected straightforward from the original graph) describe the finite sets of the edges between
users and items, items and tags, users and tags. In the real-world online social tagging systems, the graphs are very sparse.

In the aspect of the users, the tripartite graph can be projected into two bipartite graphs, which can be denoted as the user-item bipartite graph and the user-tag bipartite graph:
\begin{equation}
    G_{ur} = (U,R,E_{ur}),
    \label{eq:2}
\end{equation}
\begin{equation}
    G_{ut} = (U,T,E_{ut}).
    \label{eq:3}
\end{equation}
Thus, the users can be characterized by the resource usage information (e.g. items) and the annotation action information (e.g. tags).
The projected bipartite graphs are illustrated in Fig.2(a) and Fig.2(b), respectively.

\begin{figure}[htbp]
    \centering
    \subfigure[the user-item bipartite graph.]{
        \includegraphics[width=0.48\columnwidth]{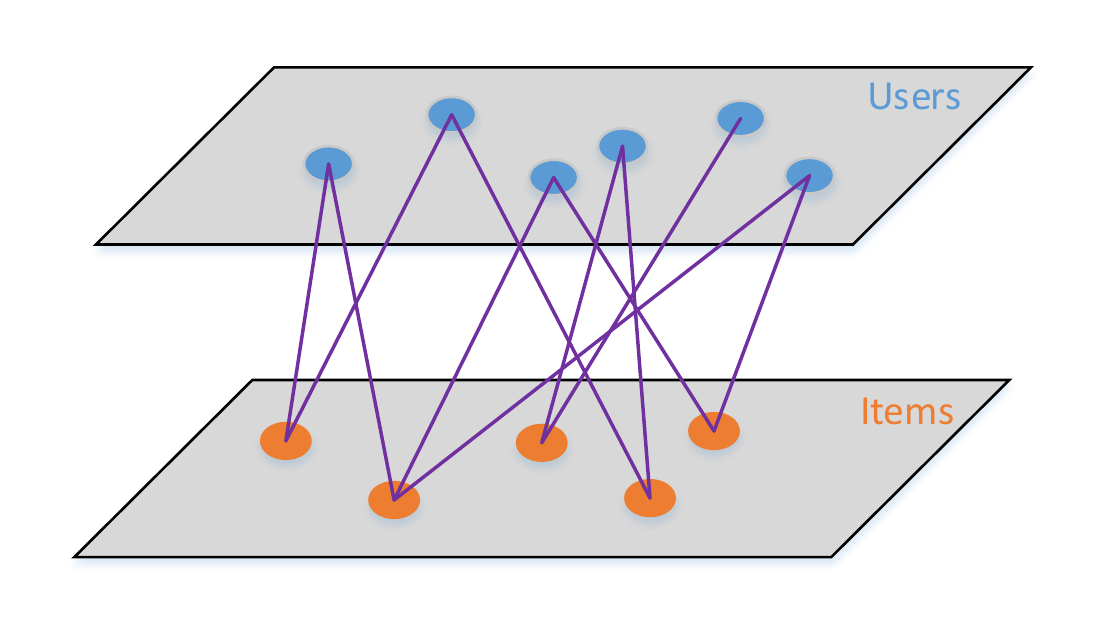}}
    \subfigure[the user-tag bipartite graph.]{
        \includegraphics[width=0.465\columnwidth]{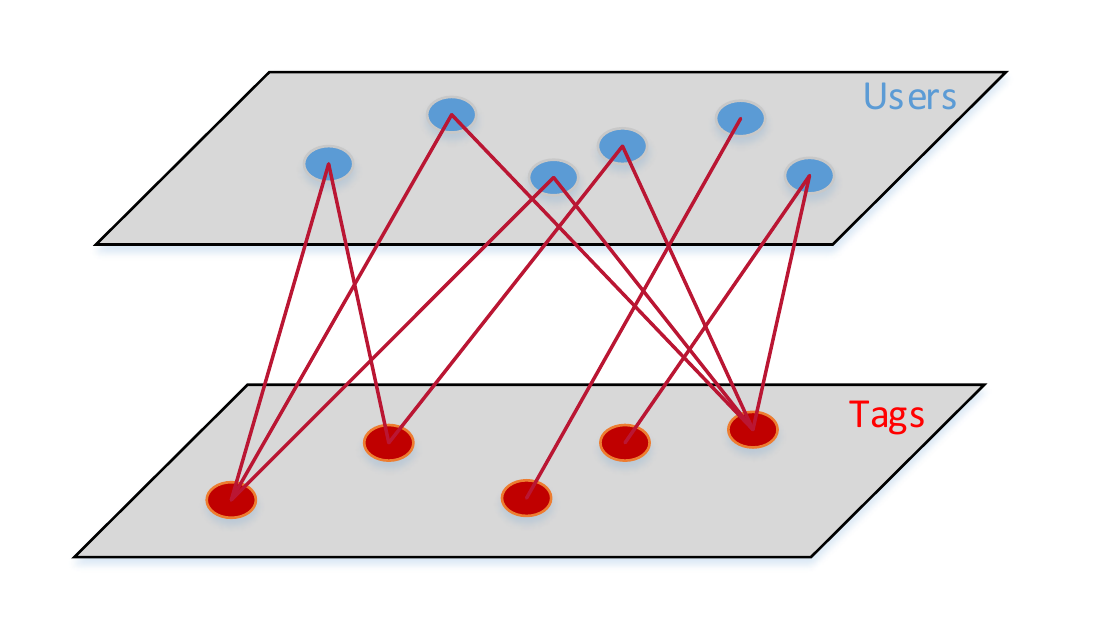}}
    \caption{The projected bipartite graphs from the tripartite graph.}
\end{figure}

In other words, the interests of a user can be represented as two vectors: the user-item vector and the user-tag vector.
The user-item vector can be denoted as:
\begin{equation}
    \vv{V_{u_{i}}^{R}} = (e_{u_{i}}^{r_{1}}, e_{u_{i}}^{r_{2}}, ... , e_{u_{i}}^{r_{N_{R}}})
    \label{eq:4}
\end{equation}
where $\vv{V_{u_{i}}^{R}}$ represents the characteristic of user $i$ in the aspect of the items, $N_{R}$ represents the total number of the items and the $e_{u_{i}}^{r_{j}}$ is as follow:
\begin{equation}
    e_{u_{i}}^{r_{j}} =
    \begin{cases}
    1,   &   \text{user $i$ tagged item $j$.} \\
    0,   &   \text{otherwise.} \\
    \end{cases}
    \label{eq:5}
\end{equation}
Similarly, the user-tag vector can be denoted as:
\begin{equation}
    \vv{V_{u_{i}}^{T}} = (e_{u_{i}}^{t_{1}}, e_{u_{i}}^{t_{2}}, ... , e_{u_{i}}^{t_{N_{T}}})
    \label{eq:6}
\end{equation}
where $\vv{V_{u_{i}}^{T}}$ represents the characteristic of user $i$ in the aspect of the tags, $N_{T}$ represents the total number of the tags and the $e_{u_{i}}^{t_{j}}$ is as follow:
\begin{equation}
    e_{u_{i}}^{t_{j}} =
    \begin{cases}
    1,   &   \text{user $i$ used tag $j$.} \\
    0,   &   \text{otherwise.} \\
    \end{cases}
    \label{eq:7}
\end{equation}

The item nodes and the tag nodes can be modeled in the same way above. Although the FCUM is general purpose for arbitrary tripartite graphs,
in this paper, we just concentrate on the characteristics of the users and recommend the items to the users. This model can be easily extended
and the framework of the fast recommendation algorithm for social tagging systems is illustrated in Fig.3.

\begin{figure}[htbp]
\centering
    \includegraphics[width=\columnwidth]{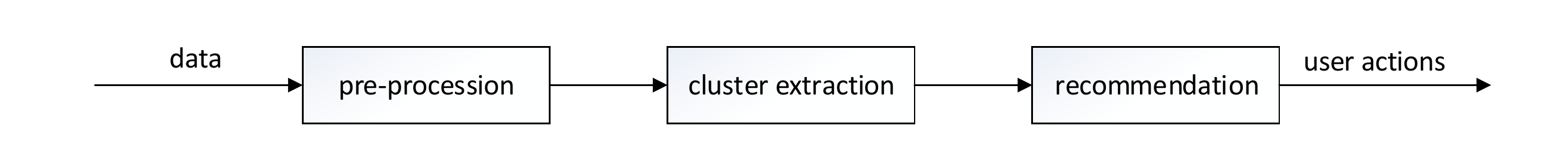}
    \caption{The framework of our fast recommendation algorithm for social tagging systems.}
\end{figure}

\subsection{Similarity}
Memory-based collaborative filtering techniques (e.g. UCF) and clustering algorithms rely on the notion of similarity between pairs of users \cite{Admavicius.(2005),Bernardes.(2015),Nanopoulos(2009)}.
The similarity between user $i$ and user $j$ can be represented through many kinds of similarity measures, such as Pearson correlation coefficient \cite{Admavicius.(2005)}, cosine similarity \cite{Admavicius.(2005),Bernardes.(2015),Nanopoulos(2009)} and Euclidean distance \cite{Nanopoulos(2009)}.
In \cite{Nanopoulos(2009),Francois(2007)}, the influence of the high dimensional and sparse data to the Euclidean distances has been studied.
They found that when the data is high dimensional and sparse, its Euclidean distances seem to concentrate and all the Euclidean distances
between pairs of data elements seem to be very similar. Moreover, the Pearson correlation coefficient is more suitable for scoring systems rather than tagging systems.
Herein, the cosine similarity is used and described as:
\begin{equation}
    cos\textrm{-}sim(u_i,u_j) = \frac{\vv{V_{u_i}} \bm\cdot \vv{V_{u_j}}}{\Vert \vv{V_{u_i}} \Vert \Vert \vv{V_{u_j}} \Vert}.
    \label{eq:8}
\end{equation}

Both the resource usages and annotation actions can reveal the interests of the users in the social tagging systems.
In this paper, we jointly take into account the resource usages and annotation actions to calculate the similarity of two users.
It is computed as:
\begin{equation}
    sim(u_i,u_j) = \beta \frac{\vv{V_{u_i}^R} \bm\cdot \vv{V_{u_j}^R}}{\Vert \vv{V_{u_i}^R} \Vert \Vert \vv{V_{u_j}^R} \Vert} + (1 - \beta) \frac{\vv{V_{u_i}^T} \bm\cdot \vv{V_{u_j}^T}}{\Vert \vv{V_{u_i}^T} \Vert \Vert \vv{V_{u_j}^T} \Vert}
    \label{eq:9}
\end{equation}
where $\beta$ is a parameter ranging from 0 to 1. Therefore, we evaluate
the similarity of two users by means of improved cosine similarity in Eq.(\ref{eq:9}).
In the FCUM, the value of $\beta$ is set as 0.5 due to the fact that
these two types of cosine similarities follow a similar distribution in the Delicious dataset.


\subsection{Cluster Extraction}
The most crucial part of the FCUM is cluster extraction. In order to accelerate the following UCF,
we need to extract useful information from massive data in the social tagging applications. In this subsection,
we will introduce the extracting procedure in details and give out its pseudo-code.

Clustering algorithms are involved in many different domains, which are used to detect the clusters in social tagging systems in \cite{Ramage(2009),Lu.(2009)},
to handle image segmentation in \cite{Isa(2009)}, to extract the social dimension in \cite{Tang.(2010)}, etc. In the FCUM, a coarse clustering algorithm,
whose similarity measure is based on Eq.(\ref{eq:9}), is used to extract useful information to accelerate the recommendation procedure without accuracy loss.
It partitions the users in the social tagging applications into non-overlapping clusters, consecutively,
the items are also divided into corresponding overlapping clusters. Figure 4 illustrates a visual view of the result.
\begin{figure}[htbp]
\centering
    \includegraphics[width=9cm,height=5cm]{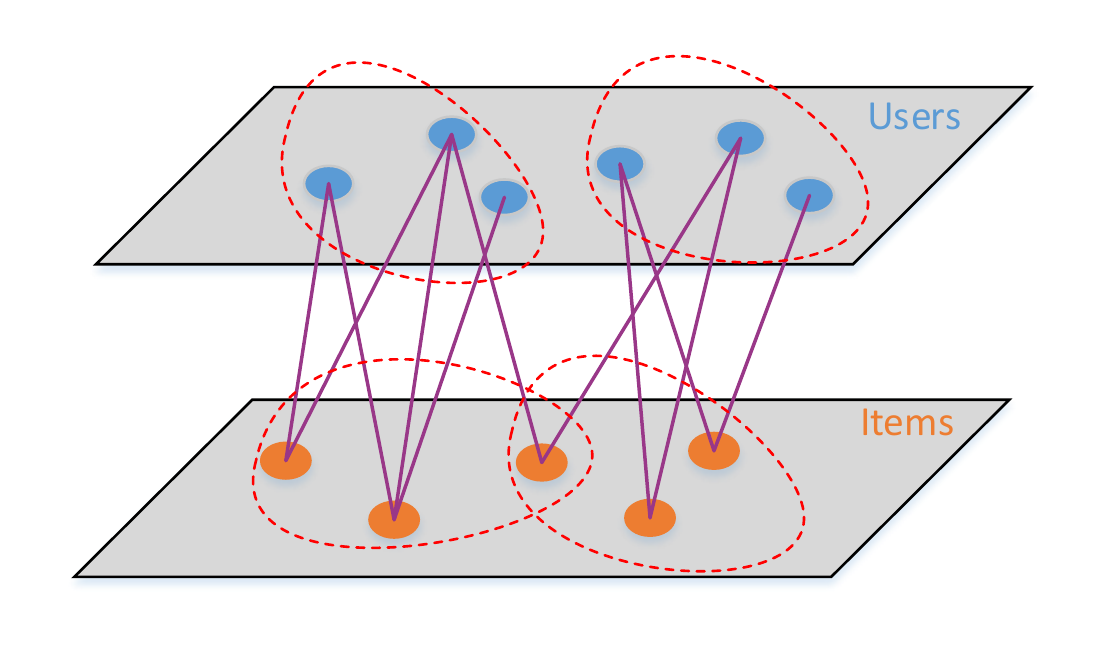}
    \caption{The result description of the coarse clustering procedure.}
\end{figure}

For the coarse clustering algorithm, its steps are similar to those in the K-means approach. The purpose of the cluster extraction procedure is
not to obtain convergent user/item clusters of the social tagging systems but to extract \emph{only} part of the user/item information for
accelerating the following UCF. Therefore, it is unnecessary to make the algorithm iterate to convergence. More concretely,
look back into the K-means approach~\cite{Jain(1999),Kanungo(2002),Kriegel(2009)}, the first step is that distributing the nodes to arbitrary clusters,
and the second one, which will be iterated many times to a convergent result, is to calculate the centroid of each cluster and redistribute each node
to a new cluster based on the similarity of the node and the centroid. In the present work, for the second step, it only needs to set a \emph{low}
iteration times to coarsely cluster these users and items. The experimental results (more details presented in Sec. 4) show that this operation
(even set the iteration times as 2) rarely affect the accuracy evaluation indicators of the recommender system. In this way, we accelerate the procedure
of cluster extraction.

To keep our description of K-means-like approach much more self-contained, we show a comprehensive operating process.
Let $K_c$ be the number of the user clusters, and $C_j^U$ $(1\leq j \leq K_c)$ represents the user cluster whose index is $j$.
It has been known that each user is characterized as two vectors in Eq.(\ref{eq:4}) and Eq.(\ref{eq:6}). The centroid
of each user cluster includes two parts:
\begin{equation}
    \vv{cent_{C_j^U}^R} = \frac{1}{N_{C_j^U}}\sum_{u_i \in C_j^U} \vv{V_{u_i}^R},
    \label{eq:10}
\end{equation}
\begin{equation}
    \vv{cent_{C_j^U}^T} = \frac{1}{N_{C_j^U}}\sum_{u_i \in C_j^U} \vv{V_{u_i}^T},
    \label{eq:11}
\end{equation}
where $N_{C_j^U}$ stands for the number of users in the user cluster $j$.
Then, the similarity between the user and corresponding centroid is computed as:
\begin{equation}
    sim(u_i,C_j^U) = \gamma \frac{\vv{V_{u_i}^R} \bm\cdot \vv{cent_{C_j^U}^R}}{\Vert \vv{V_{u_i}^R} \Vert \Vert \vv{cent_{C_j^U}^R} \Vert} + (1-\gamma)\frac{\vv{V_{u_i}^T} \bm\cdot \vv{cent_{C_j^U}^T}}{\Vert \vv{V_{u_i}^T} \Vert \Vert \vv{cent_{C_j^U}^T} \Vert,}
    \label{eq:12}
\end{equation}
where $\gamma$ is a parameter ranging from 0 to 1. Due to the fact that the $\frac{\vv{V_{u_i}^R} \bm\cdot \vv{cent_{C_j^U}^R}}{\Vert \vv{V_{u_i}^R} \Vert \Vert \vv{cent_{C_j^U}^R} \Vert}$ and the $\frac{\vv{V_{u_i}^T} \bm\cdot \vv{cent_{C_j^U}^T}}{\Vert \vv{V_{u_i}^T} \Vert \Vert \vv{cent_{C_j^U}^T} \Vert}$ follow similar distributions, $\gamma$ is set as 0.5.
On this basis, we extract the non-overlapping user clusters and corresponding overlapping item clusters (i.e., each
item cluster is constrained by the user-item bipartite graph, which brings that some of items overlaps in different clusters).

Algorithm 1 illustrates the pseudo-code. It is necessary to point out that the time complexity of Algorithm 1 is $O(T(K_c+N_U)(N_R+N_T))$
where $T$ stands for the number of iteration times, $K_c$ represents the number of user clusters, and $N_U$, $N_R$, $N_T$ are the
number of users, items and tags. In the following, the procedure of UCF is only conducted to recommend items
in $C_j^R$ (item cluster) to the associated users in $C_j^U$ (user cluster), which extremely accelerate the recommendation process.

\floatname{algorithm}{Algorithm}
\renewcommand{\algorithmicrequire}{\textbf{Input:}}
\renewcommand{\algorithmicensure}{\textbf{Output:}}
\begin{algorithm}[!htb]
    \caption{user and item clusters extraction}
    \begin{algorithmic}[1]
        \Require the user-item bipartite graph $G_{ur}$; the user-tag bipartite graph $G_{ut}$; the number of user clusters $K_c$; the number of iteration times $iterTime$;
        \Ensure the non-overlapping user clusters $C_j^U$; the corresponding overlapping item clusters $C_j^R$;
        \State assign each user to a random cluster
        \State $iterTime \gets 2$
        \While {$iterTime > 0$}
            \State calculate the centroids of the clusters
            \For {$j = 1 \to K_c$} \State $temp\textrm{-}C_j^U = \varnothing$  \EndFor
            \For {each user $u_i$}
                 \For {$j = 1 \to K_c$}
                    \State calculate the $sim(u_i,C_j^U)$
                 \EndFor
                 \State find the index $j$ of the cluster which maximizes $sim(u_i,C_j^U)$
                 \State $temp\textrm{-}C_j^U$ = $temp\textrm{-}C_j^U \cup \{u_i\}$
            \EndFor
            \For {$j = 1 \to K_c$} \State $C_j^U = temp\textrm{-}C_j^U$  \EndFor
            \State $iterTime = iterTime - 1$
        \EndWhile
        \For {$j = 1 \to K_c$}
            \State $C_j^R = \varnothing$
            \For {each user $u_i$ in $C_j^U$}
                \For {each resource usage $u_i\textrm{-}r_k$ in $G_{ur}$}
                    \State $C_j^R = C_j^R \cup \{r_k\}$
                \EndFor
            \EndFor
        \EndFor
    \end{algorithmic}
\end{algorithm}

\subsection{User Based Collaborative Filtering}
The UCF has been used to recommend movies, songs, jobs, books and other products in the e-commercial systems,
online social systems and other types of online applications~\cite{Admavicius.(2005),Aiolli.(2013),Bernardes.(2015)}.
In this procedure, each user obtains an item ranklist, in which he likely tends to store, tag and share the top items.
To rank the items, the score function is denoted as follows:
\begin{equation}
    score(u_i,r_k | u_i \in C_j^U) =
    \begin{cases}
        \sum_{u_s \in C_j^U}f[sim(u_i,u_s)|r_k],    &       e_{u_i}^{r_k} == 0  \\
        -1,                                                      &       otherwise           \\
    \end{cases}
    \label{eq:13}
\end{equation}
\begin{equation}
    f[sim(u_i,u_s)|r_k] =
    \begin{cases}
        sim(u_i,u_s),   &   e_{u_s}^{r_k} == 1  \\
        0,              &   otherwise           \\
    \end{cases}
    \label{eq:14}
\end{equation}
After the calculation, we sort the scores and get the ranklist for every user. Because a higher score
represents that a user is more likely to store, tag and share the item, we recommend the item with
a higher user specific score to the user.

In addition, the time complexity of the UCF is $O(N_U(N_UN_R+N_T))$, where $N_U$, $N_R$, and $N_T$
represent the number of users, items and tags, respectively. However, In the FCUM, the time cost can be represented as
$O(\sum_{j=1}^{K_c}N_{C_j^U}(N_{C_j^U}N_{C_j^R}+N_{C_j^T}))$, where $N_{C_j^U}$ stands for the number of users in the $j$th user cluster,
$N_{C_j^R}$ denotes the number of items in the $j$th item cluster, and $N_{C_j^T}$ is the number of tags which are associated with the users in the $j$th user cluster.
Obviously, due to the fact that $\sum_{j=1}^{K_c}N_{C_j^U} == N_U$, the time cost of the proposed algorithm is less than that of the ordinary one
even including the time cost of the coarse clustering procedure.

\subsection{Extending model}
In this paper, we propose the foundation of the FCUM that only uses the information from resource usages and annotation actions to construct the model.
However, in the real-world social tagging applications, we can get information from user profiles, explicit user relationships,
as well as implicit relationships among the users and items such as the situations that two users both click one link but do not tag it
or the IP addresses of two users are located in the same city. These features are able to be represented in the vector forms
like Eq.(\ref{eq:4}) and Eq.(\ref{eq:6}), then add them to the FCUM. Thus, the similarities among users can be extended as:
\begin{equation}
    sim(u_i,u_j)= \sum_{k=1}^{N_f}\beta_k\frac{\vv{V_{u_i}^{f_k}} \bm\cdot \vv{V_{u_j}^{f_k}}}{\Vert \vv{V_{u_i}^{f_k}} \Vert \Vert  \vv{V_{u_j}^{f_k}} \Vert}
    \label{eq:15}
\end{equation}
\begin{equation}
    \sum_{k=1}^{N_f}\beta_k = 1,
    \label{eq:16}
\end{equation}
where $N_f$ is the number of feature vectors, $\vv{V_{u_i}^{f_k}}$ is the $k$-th feature vector of user $u_i$, and $\beta_k$ is the parameter ranging from 0 to 1.
Similarly, in the cluster extraction procedure introduced in Sec. $2.2$, the centroid of each cluster can be
represented as $N_f$ vectors. Thus, it is able to construct a hybrid FCUM\cite{Jin(2011)}.

\section{Experimental Design}

\subsection{Evaluation Indicators}
To evaluate whether the recommended items meet the users' interests based on the FCUM, we divide the available
dataset into training and testing subsets according to the timestamps. And, three common evaluation indicators,
recall, precision and F1-score, are described respectively \cite{Bernardes.(2015)}:

\begin{equation}
    recall@k = \frac{1}{N_U} \sum_{i=1}^{N_U} \frac{\Vert R_{u_i}^k \cap T_{u_i} \Vert}{\Vert T_{u_i} \Vert}
    \label{eq:17}
\end{equation}
\begin{equation}
    precision@k = \frac{1}{N_U} \sum_{i=1}^{N_U} \frac{\Vert R_{u_i}^k \cap T_{u_i} \Vert}{k}
    \label{eq:18}
\end{equation}
\begin{equation}
    f_1@k = \frac{2 \times precision@k \times recall@k}{precision@k + recall@k}
    \label{eq:19}
\end{equation}
where $k$ is the ranklist (i.e., recommendation list) length, $R_{u_i}^k$ is the finite set of items $\{r_{u_i}^1, r_{u_i}^2, ..., r_{u_i}^k\}$ recommended to user $u_i$, $T_{u_i}$ is the test set for user $u_i$.
The recall describes the true positive rate, the precision is referred to the positive predictive value, and F1-score combines the recall and the precision into the harmonic mean.

\subsection{Dataset and Platform}
In the real-world social tagging systems, the tripartite graphs are usually sparse.
It suggests that many users have only few resource usages and annotation actions,
and many items are stored, tagged and shared only few times. Thus, the accuracy and time performance will be obviously influenced. Before conducting experiment,
a pre-process is done for the original dataset.

The original dataset is crawled from a real-world social tagging system, Delicious.
It contains 1867 users, 69226 URLs (items), 53388 tags, and 437595 user-URL-tag triples.
In the pre-process, we filter out the user nodes, resource nodes, tag nodes,
whose degrees are lower than a threshold by iteratively removing these nodes and edges from the graph.
For example, when the threshold equals to 5, there are 1617 users, 21983 URLs, 5301 tags and 236659
user-URL-tag triples in the filtered graph. Note that we also consider other thresholds in the
following experiments. After that, according to the timestamp, the filtered dataset is
divided into 80$\%$ training subset and 20$\%$ testing subset in terms of the user-URL-tag triples.
Moreover, in this procedure, we make sure that every user is in the testing subset. Table 1 is characterized by
the number of users, URLs, tags and user-URL-tag triples in the training set, in the testing set as well as in total.

The experimental platform is a notebook computer with 4 AMD A10-4600M APU cores and 4GB DRAM.
Its operating system is ArchLinux with a linux kernel of 4.1.5. We use g++ 5.2.0 with $-O2$ compiler optimization level.

\begin{table}[htbp]
    \centering
    \begin{tabular}{| m{0.15\columnwidth}<{\centering} | m{0.1\columnwidth}<{\centering} | m{0.1\columnwidth}<{\centering} | m{0.1\columnwidth}<{\centering} | m{0.3\columnwidth}<{\centering}|}
        \hline
         & users & URLs & tags &  user-URL-tag triples \\
        \hline
        training set & 1617 & 20338 & 5299 & 188671 \\
        \hline
        testing set & 1617 & 8055 & 4758 & 47988 \\
        \hline
        total & 1617 & 21983 & 5301 & 236659 \\
        \hline
    \end{tabular}
    \caption{The statistics of the filtered dataset(degree threshold = 5)}
\end{table}

\begin{table}[htbp]
    \centering
    \begin{tabular}{| m{0.17\columnwidth}<{\centering} | m{0.1\columnwidth}<{\centering} | m{0.1\columnwidth}<{\centering} | m{0.1\columnwidth}<{\centering} | m{0.1\columnwidth}<{\centering} | m{0.1\columnwidth}<{\centering}| m{0.12\columnwidth}<{\centering} | }
        \hline
        indicators & $n = 2$ & $n = 4$ & $n = 6$ & $n = 8$ & $n = 10$ & \textbf{UCF} \\
        \hline
        $recall@5$ & \textbf{0.11916} & 0.11676 & 0.11197 & 0.11324 & 0.11524 & 0.11146 \\
        $precision@5$ & \textbf{0.05244} & 0.05071 & 0.04799 & 0.04836 & 0.05022 & 0.0491 \\
        $F_1@5$ & \textbf{0.07283} & 0.07071 & 0.06718 & 0.06778 & 0.06995 & 0.06817 \\
        \hline
        $recall@10$ & \textbf{0.14772} & 0.14536 & 0.14341 & 0.14367 & 0.14441 & 0.14268 \\
        $precision@10$ & \textbf{0.03822} & 0.03643 & 0.0363 & 0.03618 & 0.03643 & 0.03599 \\
        $F_1@10$ & \textbf{0.06073} & 0.05825 & 0.05794 & 0.0578 & 0.05818 & 0.05748 \\
        \hline
        $recall@15$ & \textbf{0.16099} & 0.16033 & 0.15899 & 0.15825 &  0.15863 & 0.158 \\
        $precision@15$ & \textbf{0.03051} & 0.03039 & 0.02985 & 0.02964 & 0.02997 & 0.02956 \\
        $F_1@15$ & \textbf{0.0513} & 0.05109 & 0.05026 & 0.04993 & 0.05042 & 0.0498 \\
        \hline
        $recall@20$ & \textbf{0.16829} & 0.16751 & 0.16709 & 0.16673 & 0.1667 & 0.1676 \\
        $precision@20$ & \textbf{0.02557} & 0.02548 & 0.02514 & 0.02523 & 0.02536 & 0.02566 \\
        $f_1@20$ & \textbf{0.0444} & 0.04423 & 0.0437 & 0.04383 & 0.04402 & 0.04451 \\
        \hline
    \end{tabular}

    \caption{The influence of iteration time on performance.}
\end{table}

\section{Results and Discussion}
In order to achieve the extraction of non-overlapping user clusters and corresponding overlapping item clusters during a relatively short time, the iteration times is set
no more than 10. Moreover, the degree threshold mentioned in Sec. $3.2$ equals to 5 in terms of the sparseness of the graph.
Firstly, we explore the influence of the iteration time on the the evaluation indicators of the recommender system. In the cluster
extraction procedure, the average number of users in each cluster is 90, which roughly fixes initial 18 clusters. Note that the initially
random allocation of users to each cluster is insensitive to the experimental results (see in Fig.6). For iteration number from n=2 to n=10 (step length 2),
three evaluation indicators of the recommender system are computed restricted to ranklist length and in the comparison of UCF,
which are illustrated in Tab. 2 and in Fig.5.

\begin{figure}[htbp]
    \centering
    \subfigure[recall]{
        \includegraphics[width=0.48\columnwidth,height=4.7cm]{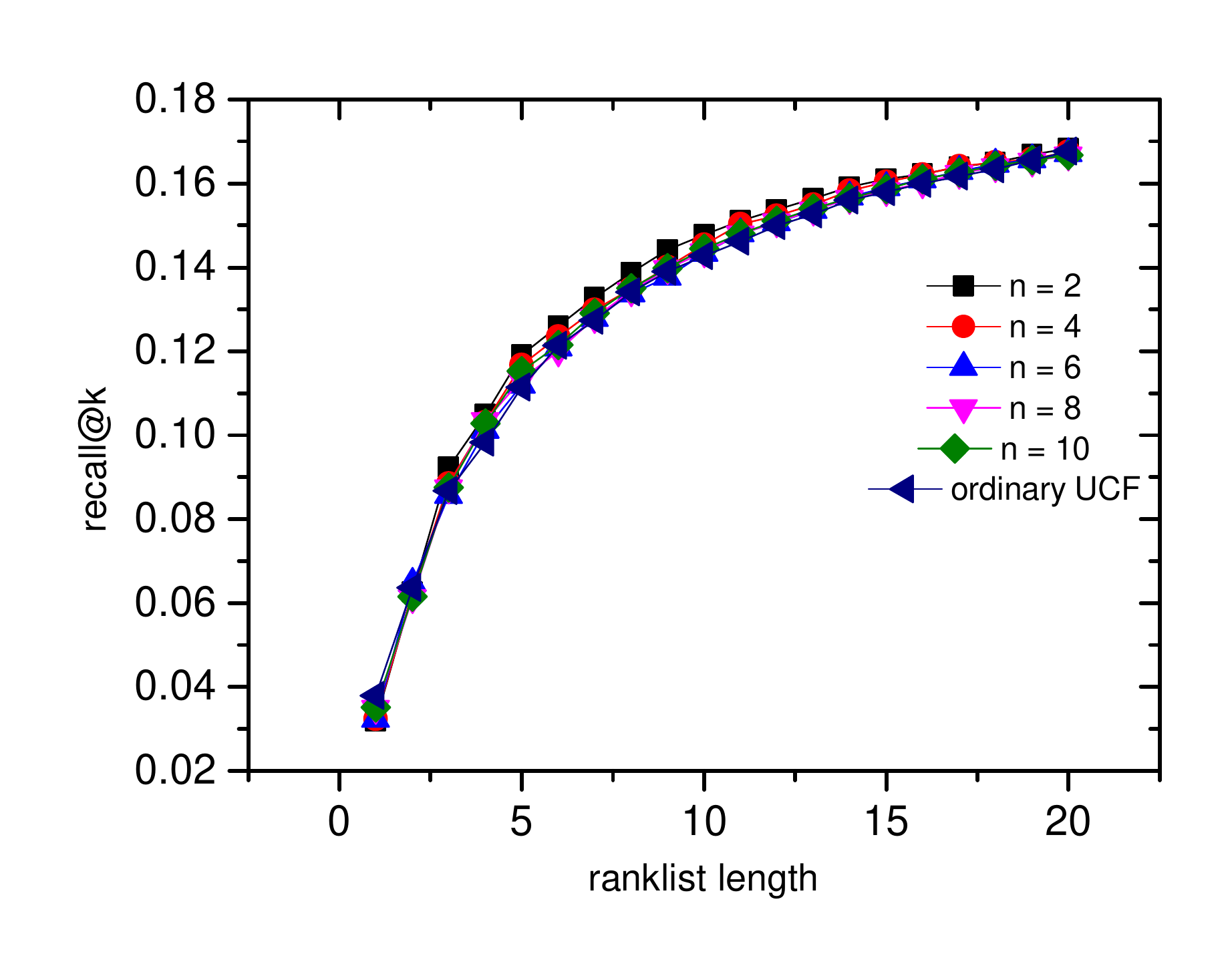}}
    \subfigure[precision]{
        \includegraphics[width=0.48\columnwidth,height=4.7cm]{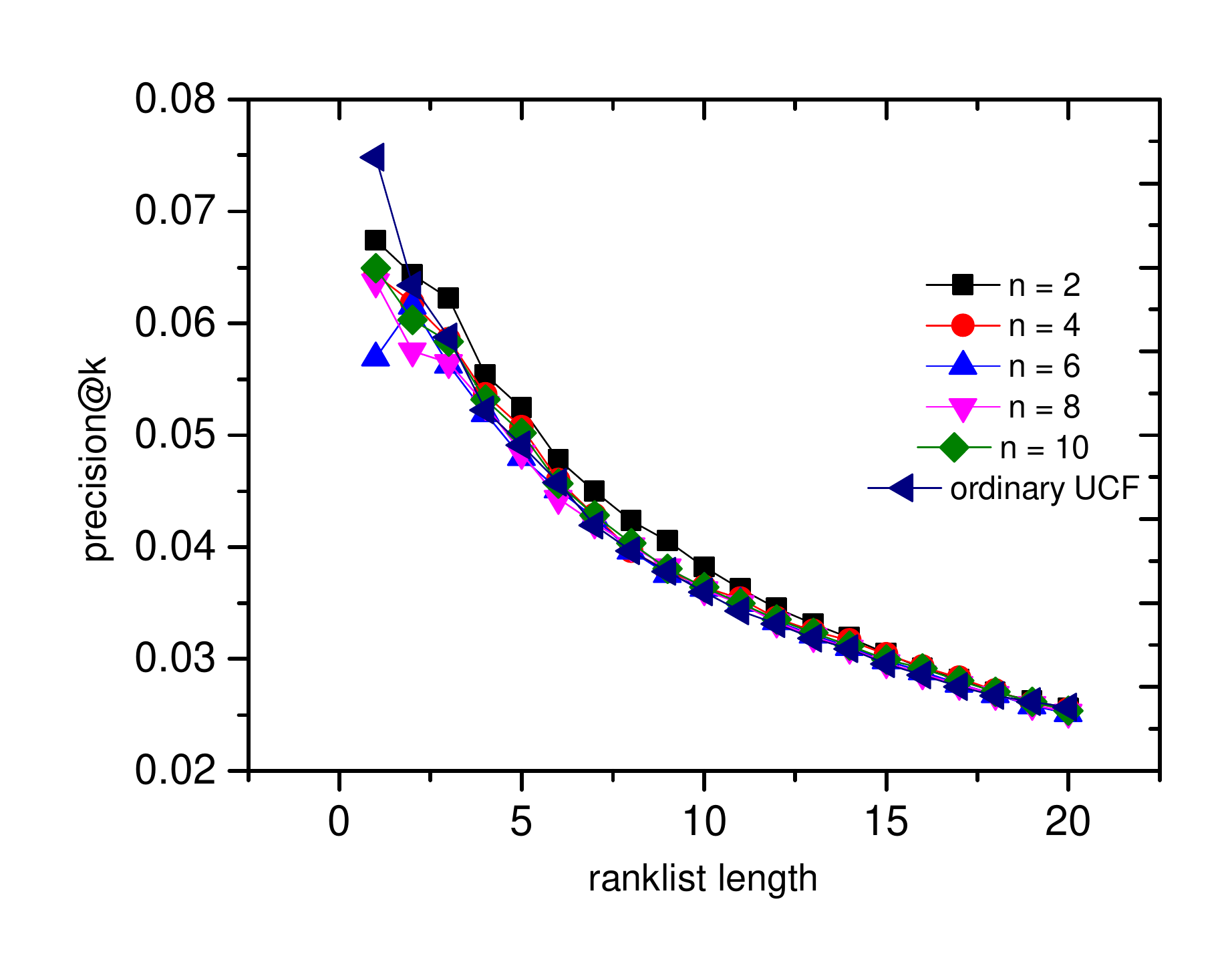}}
    \subfigure[$f_1$]{
        \includegraphics[width=0.48\columnwidth,height=4.7cm]{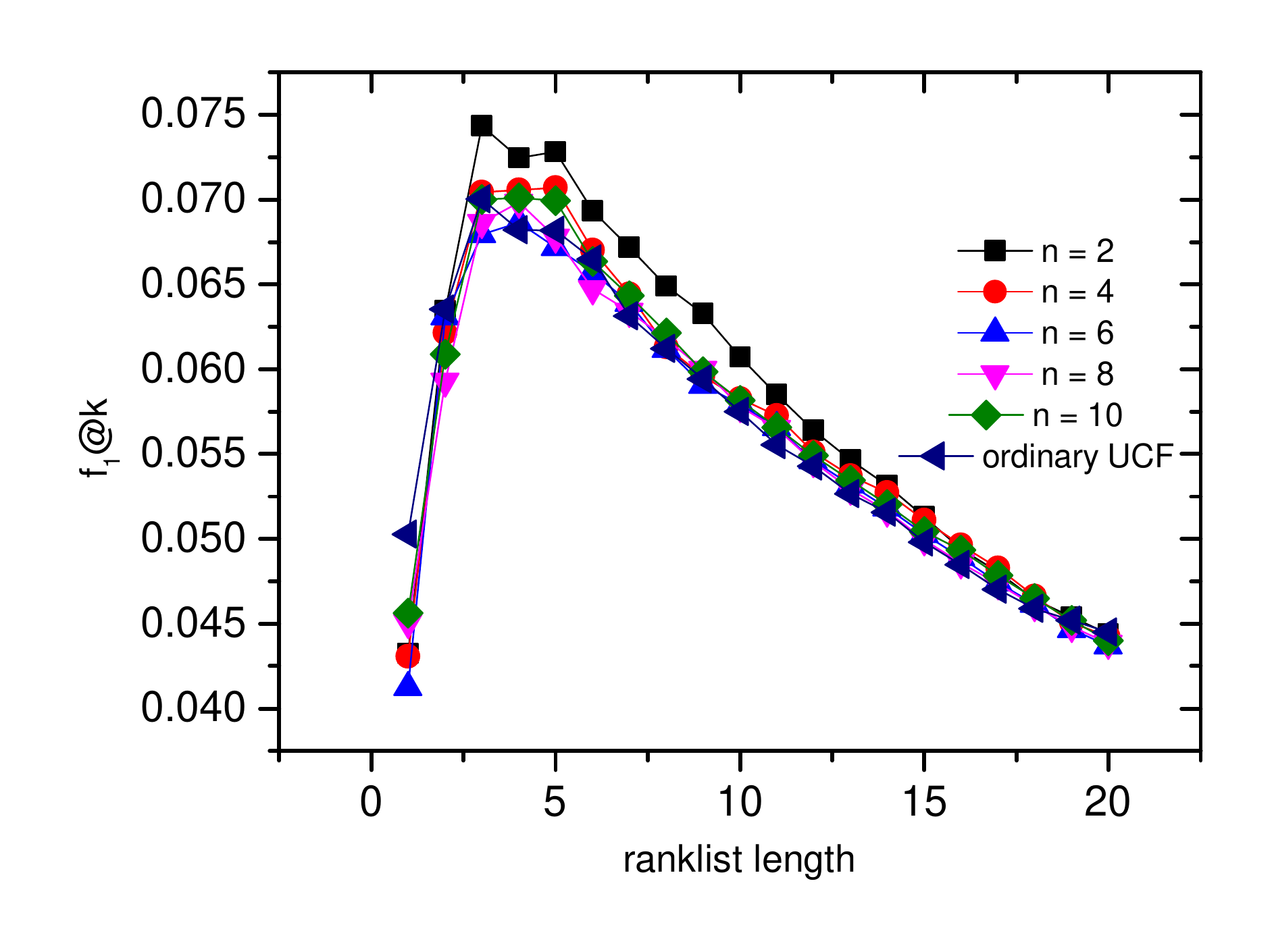}}
    \subfigure[time cost@20]{
        \includegraphics[width=0.48\columnwidth,height=4.67cm]{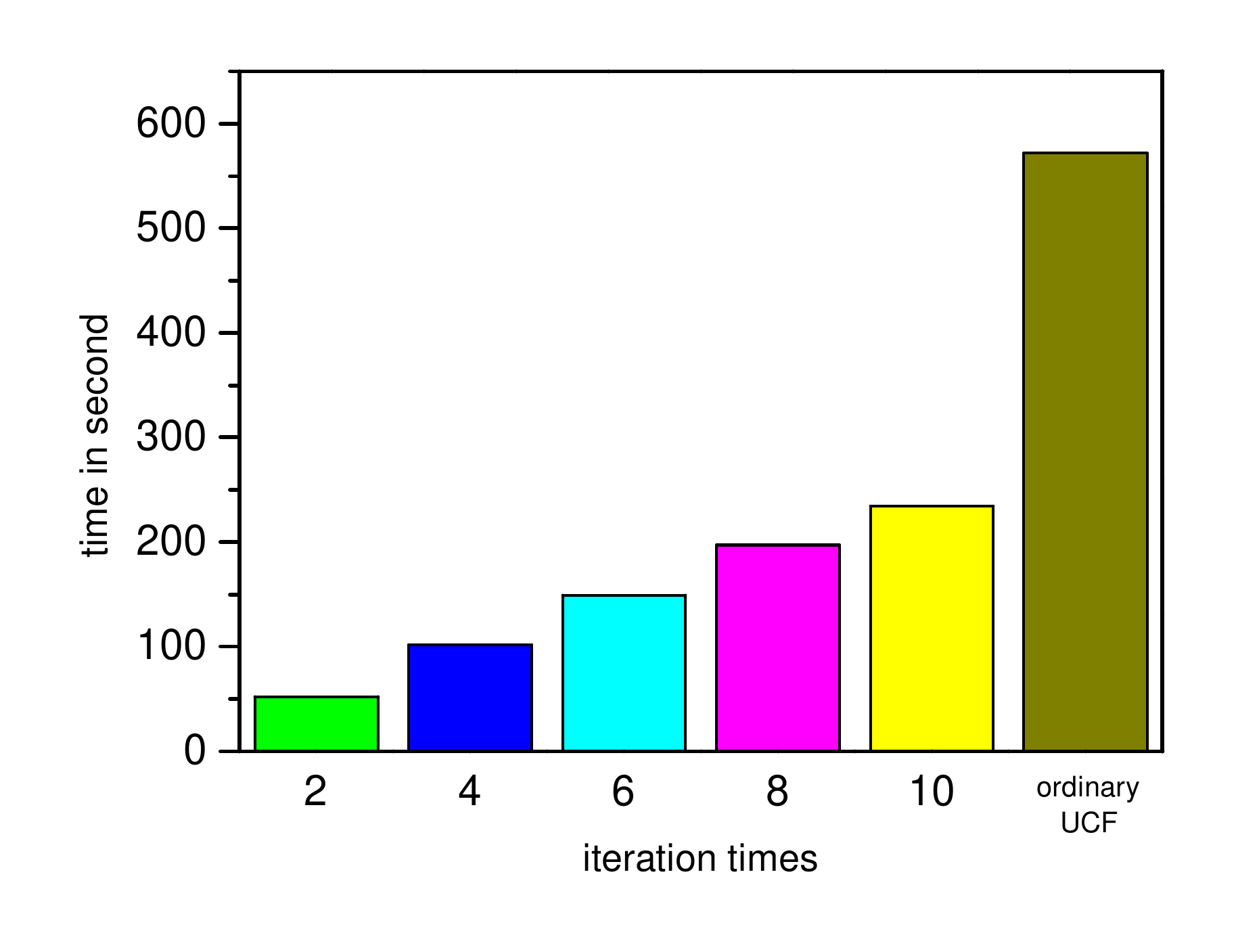}}
    \caption{(color online) The influence of iteration time on performance as a function of ranklist length.}
\end{figure}

More concretely, in Tab. 2, it can be found that these evaluation indicators are insensitive to the iteration time $n$,
suggested by their similar values in restricted to the ranklist length, and our model doesn't weaken the accuracy of the recommendation system in a comparison of the
UCF (more details shown in Fig.5). More importantly, when the iteration time equals to 2, the statistical averages of these evaluation
indicators show that the FCUM have a relatively better accuracy (see in Fig.5(a)-(c)) when the ranklist length ranges from 3 to 19, and the time performance is significantly better than the UCF
(see in Fig.5(d)), that is, the time cost is reduced greater than 90$\%$. Furthermore, the comprehensive analysis of these evaluation indicators
in Fig.5 shows that the recall increases and precision decreases as a function of the ranklist length, which makes the F1-score be optimal
when the ranklist length is 3.

\begin{figure}[htbp]
    \centering
    \subfigure[recall]{
        \includegraphics[width=0.48\columnwidth,height=4.7cm]{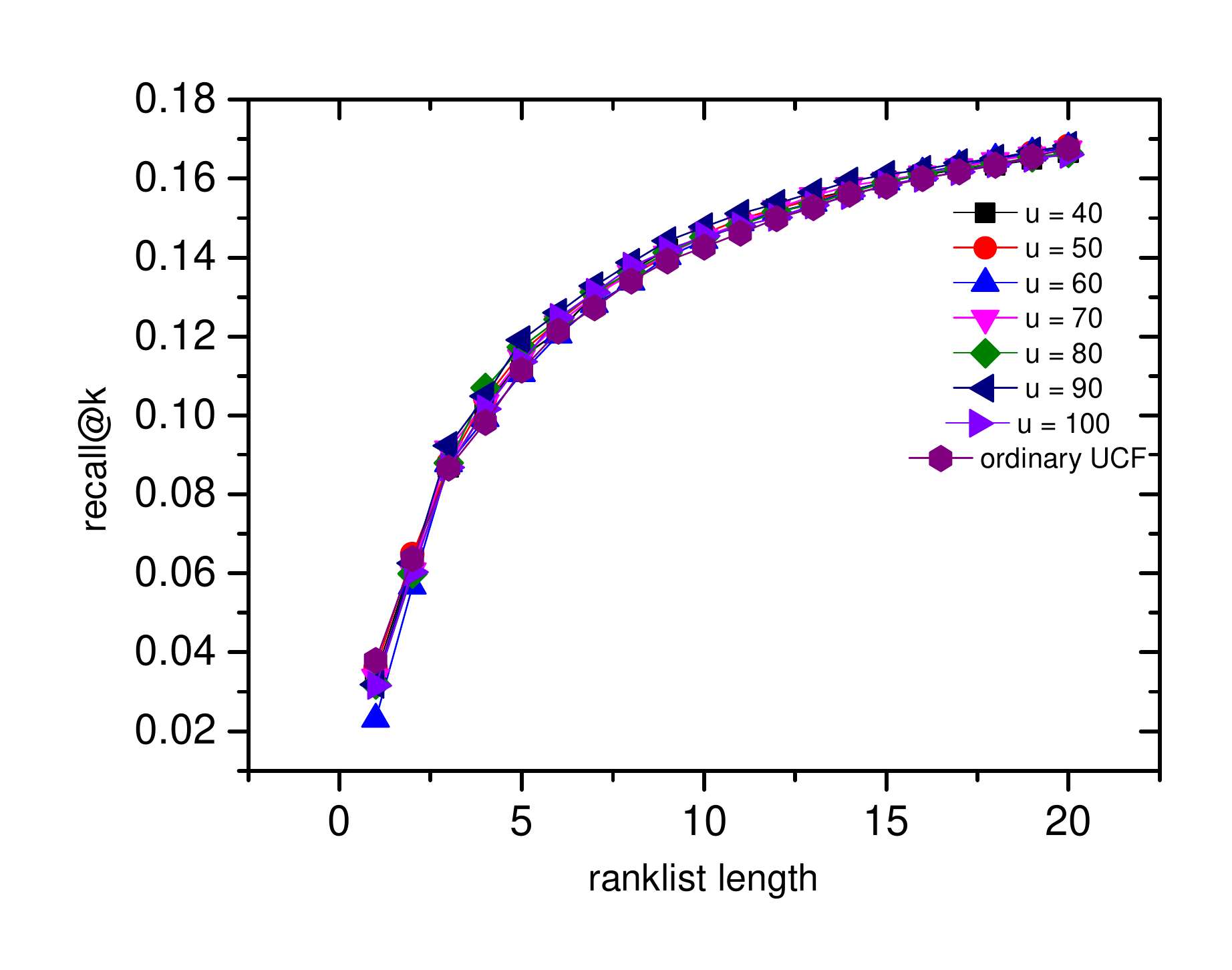}}
    \subfigure[precision]{
        \includegraphics[width=0.48\columnwidth,height=4.7cm]{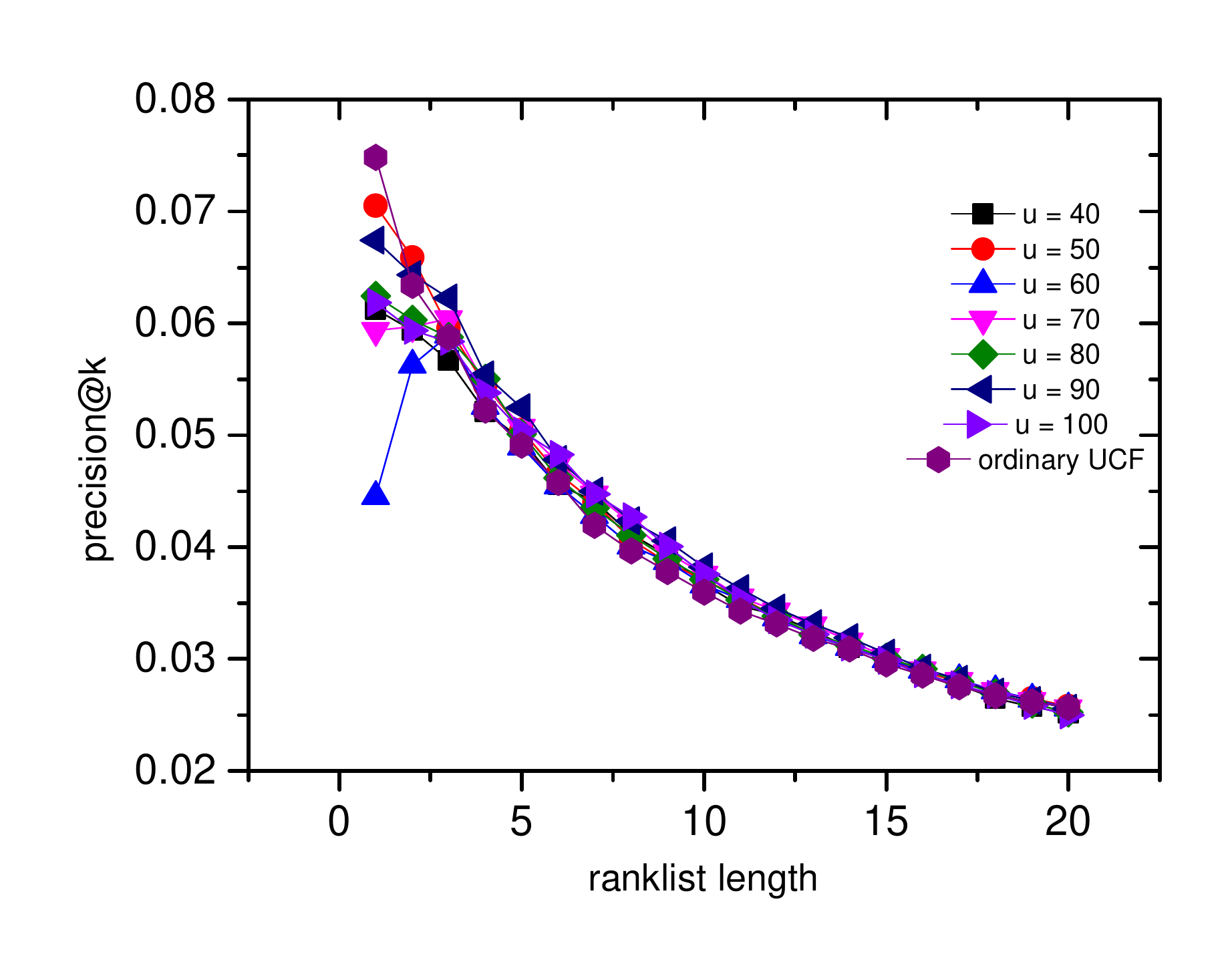}}
    \subfigure[$f_1$]{
        \includegraphics[width=0.48\columnwidth,height=4.7cm]{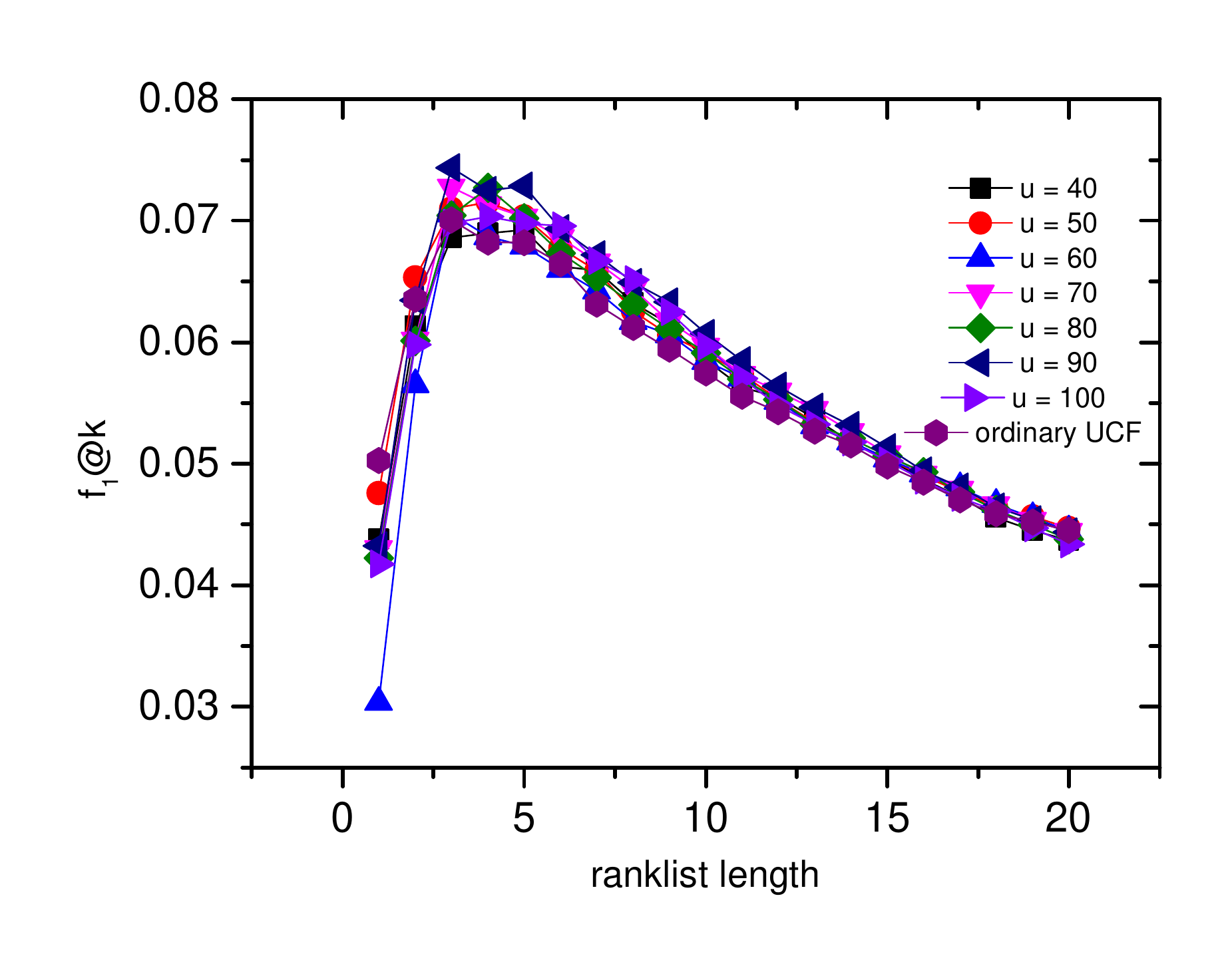}}
    \subfigure[time cost]{
        \includegraphics[width=0.48\columnwidth,height=4.54cm]{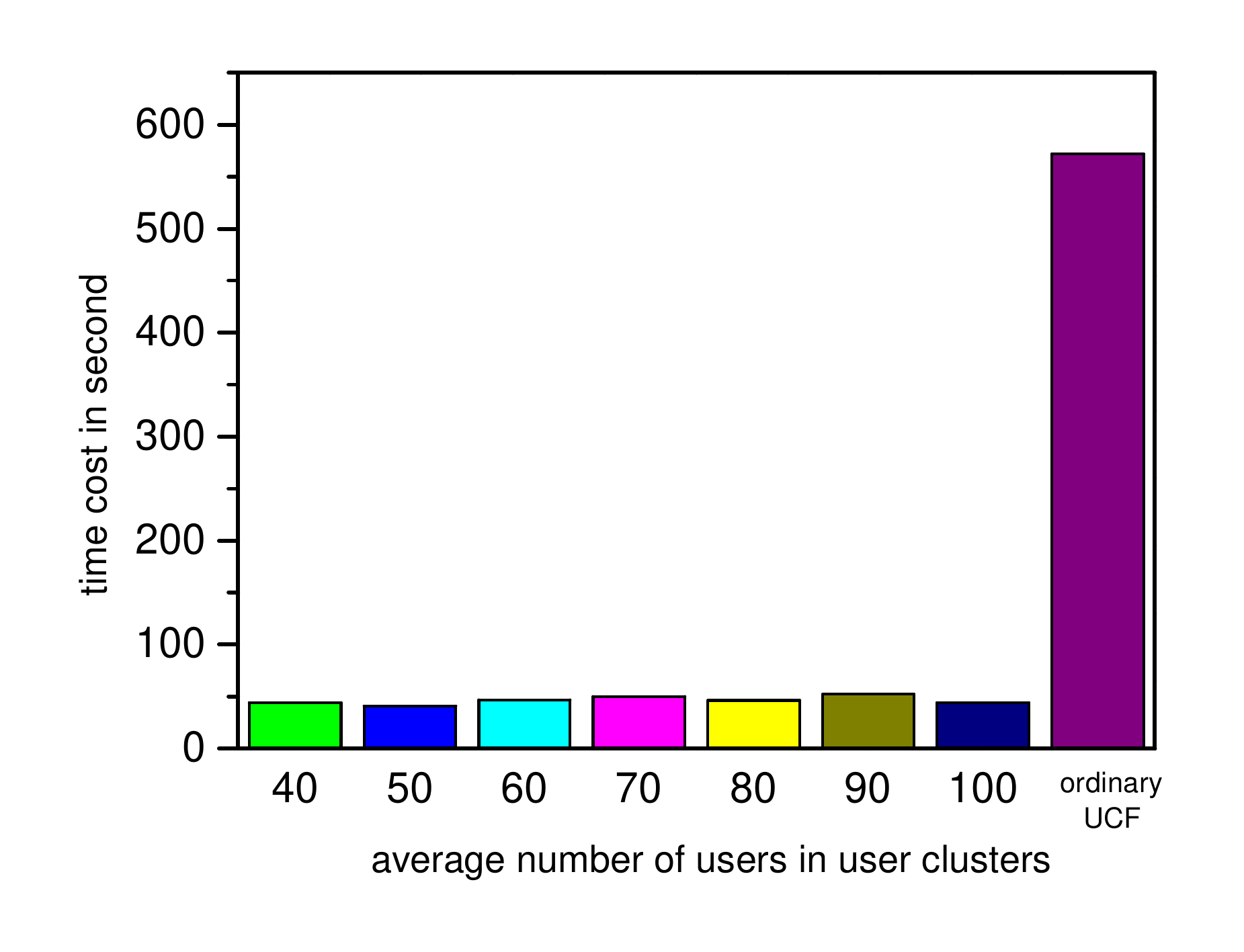}}
    \caption{(color online) The influence of the average number of users in each cluster on performance as a function of ranklist length.}
\end{figure}

As mentioned in above analysis, we set the average number of users in each cluster to 90 and obtain the initial number of clusters.
It is well known that the initial number of clusters to some extent affect the final result when cluster procedure converges. Actually,
it is unnecessary to make clustering procedure finally converge in the FCUM, thus we don't care about its convergent result, but
only concentrate on the performance of the recommender system. Nevertheless, to keep the study self-contained,
we still perform experiments that whether the average number of users in each
cluster finally affect these evaluation indicators. Herein, when the iteration time and degree threshold is 2 and 5 respectively,
we adjust the average number of users in each cluster from 40 to 100 with the step equalling 10 and independently conduct each experiment.
Figure 6 shows that the average number of users in each cluster is also a trivial factor affecting these evaluation indicators
which are suggested by their similar values in restricted to the ranklist length, as well as the lower time cost in a comparison of the UCF.
Through these experiments, we demonstrate that the FCUM not only strongly enhances the efficiency, but also relatively improves the accuracy
of the recommender system.

\begin{figure}[htbp]
    \centering
    \subfigure[recall]{
        \includegraphics[width=0.48\columnwidth,height=4.7cm]{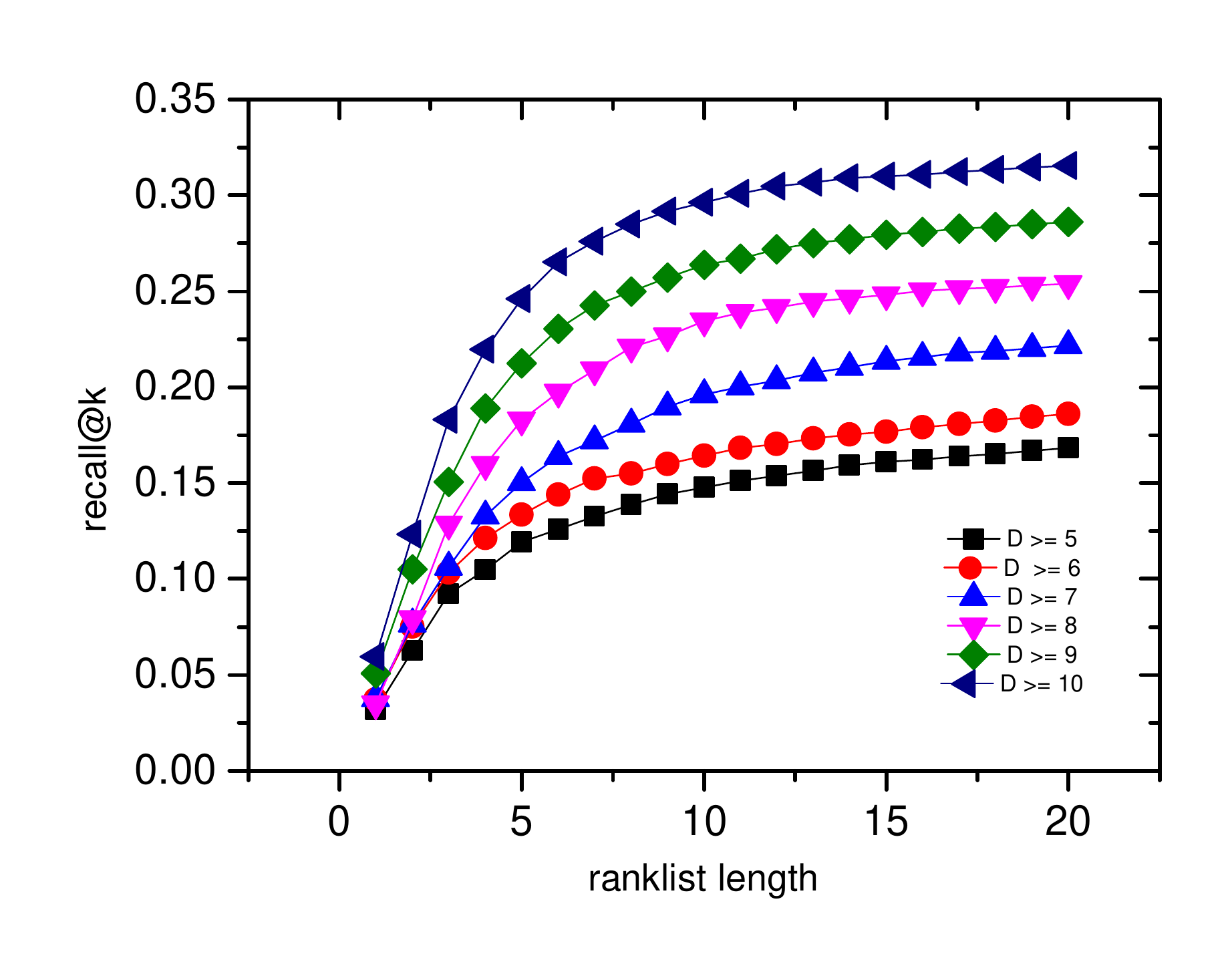}}
    \subfigure[precision]{
        \includegraphics[width=0.48\columnwidth,height=4.7cm]{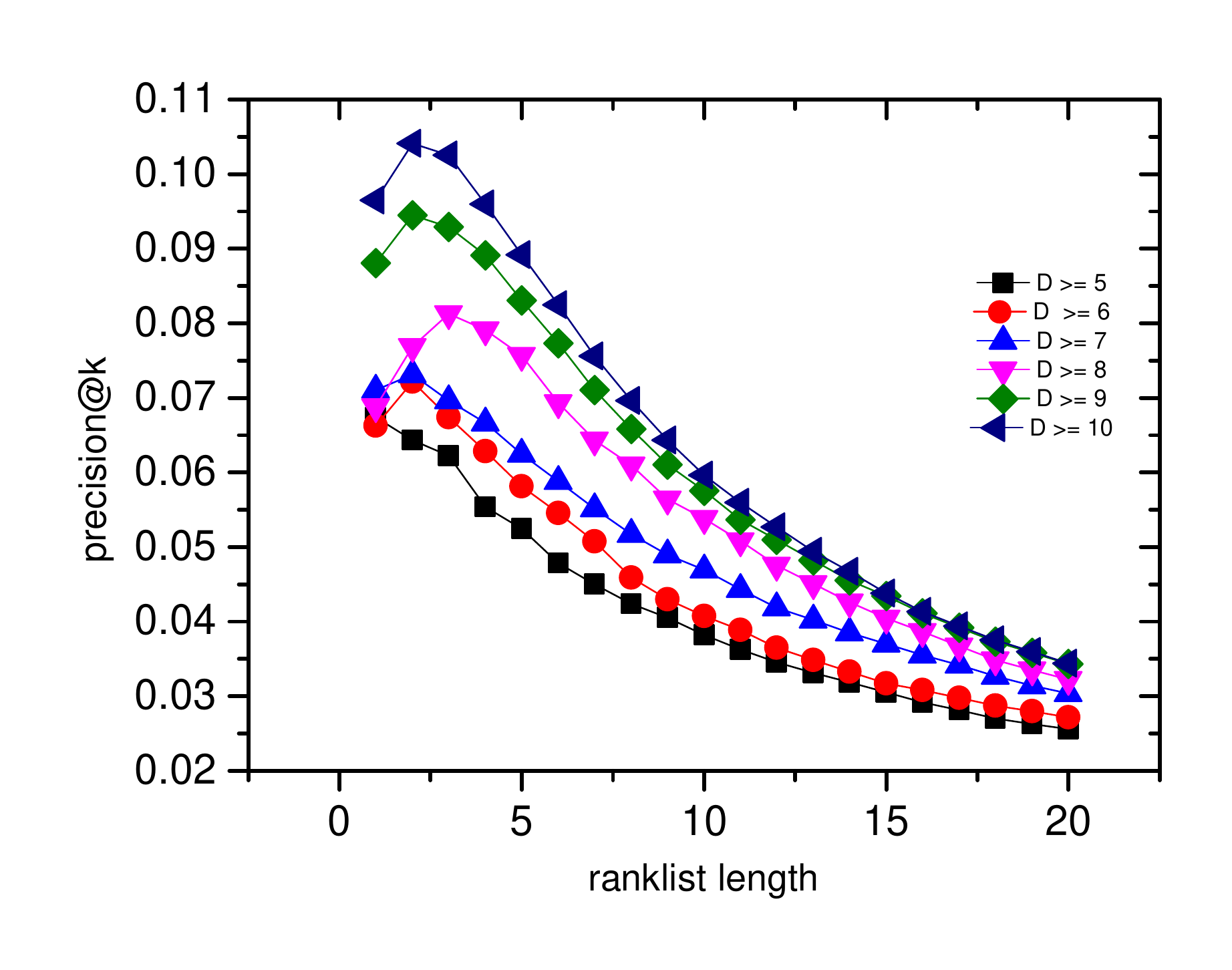}}
    \subfigure[$f_1$]{
        \includegraphics[width=0.48\columnwidth,height=4.7cm]{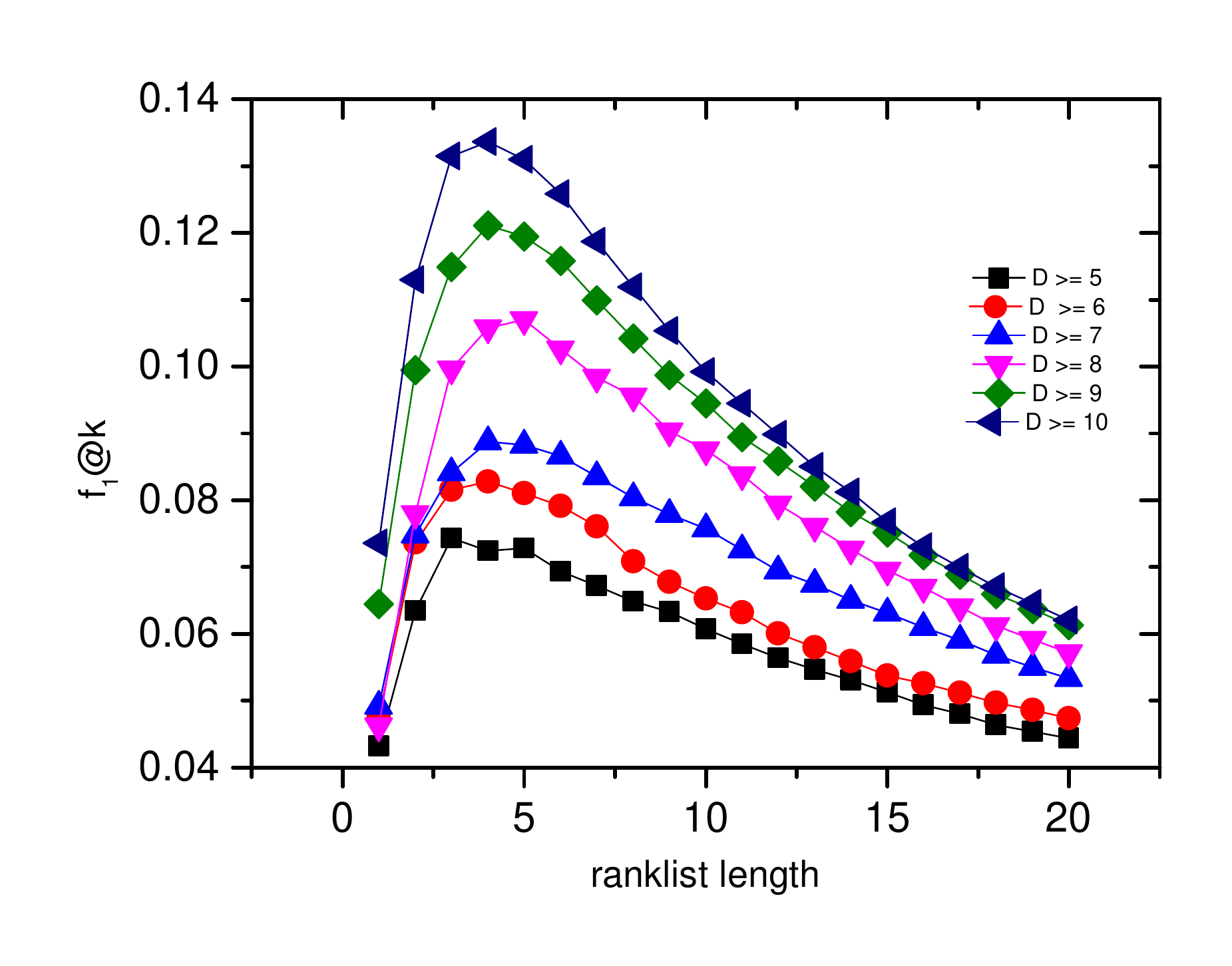}}
    \subfigure[time cost]{
        \includegraphics[width=0.48\columnwidth,height=4.65cm]{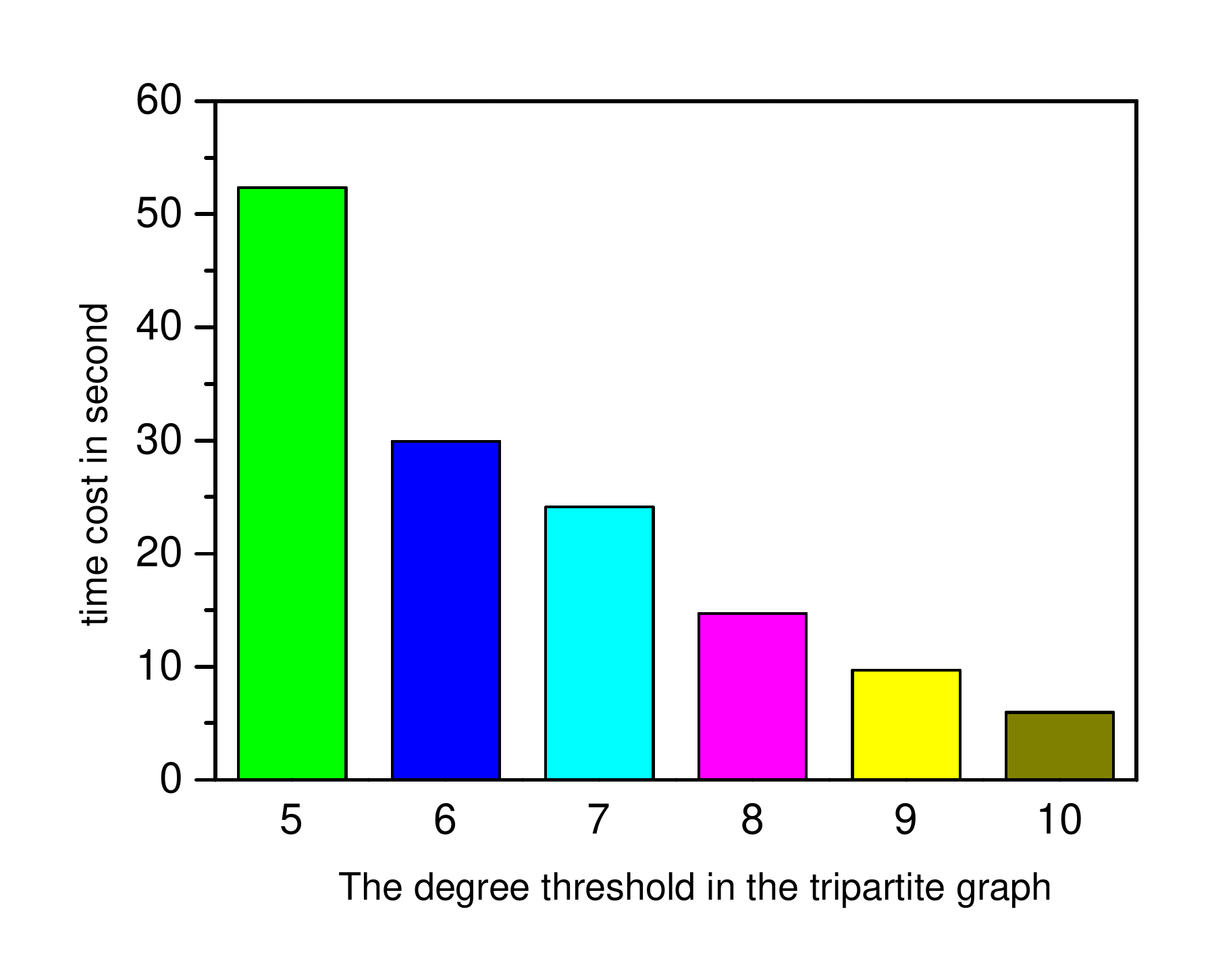}}
    \caption{(color online)the influence of the degree threshold to the performance.}
\end{figure}

In addition, in the preprocess procedure, we take account of the data sparseness of the social tagging system (or tripartite graph) and
the degree threshold of each node. Thus, It is worthy to discuss whether the different degree thresholds change the result in the FCUM. We further increase the
degree threshold from 6 to 10, and obtain these filtered tripartite graphs. According to the initial condition setting $n=2$ and $u=90$, a number of
contrastive experiments are performed based these filtered tripartite graphs, whose results are shown in Fig.7. It can be found that with the increment of the degree threshold,
the evaluation indicators of the recommendation algorithm become better. However, in these situations, only the high degree nodes are considered and it aggravates the cold start problem.
In a real-world recommender system, we need to make a trade-off among the cold start problem, the accuracy of the model and the time complexity, it's application dependent, we just give out a hint and the discussion of these issues is far beyond this paper.

\section{Conclusion}

In this paper, we propose a fast and elegant collaborative user model based on cluster extraction to recommend items to users in social tagging systems.
And the cluster extraction is insensitive to the parameters such as the iteration time and initial number of clusters. The extensive experiments
demonstrate that the recommendation algorithm based on this model behaves much more efficiently due to the fact that the time cost is dramatically reduced
greater than 90$\%$, and relatively accurately in the comparison
of UCF. Moreover, it exploits both the information of the resource usages and annotation actions, and can be extended to use more information which can be
represented as vectors extracted from the social tagging applications. As relevant issues for future work, we plan to characterize each user in the
social tagging systems not only by the information of resource usages and annotation actions, but also explore the explicit or implicit relationships between
users such as their personal attributes, friend relationships, follow relationships, which help for improving the performance of recommender system.

\section*{Acknowledge}
This work is partially supported by the National Natural Science Foundation of China (Grant Nos. 61370150 and 61433014)
and Special Project of Sichuan Youth Science and Technology Innovation Research Team (Grant No. 2013TD0006).












\end{document}